\documentclass[sigconf, nonacm]{acmart}
\usepackage[utf8]{inputenc}
\usepackage{geometry}
\usepackage{xurl}
\usepackage{hyperref}
\usepackage{graphicx}
\usepackage{tabularx}
\usepackage{multirow}
\usepackage{float}
\usepackage{enumitem} 
\usepackage{tikz}
\usepackage{microtype}
\usepackage{subcaption}
\usetikzlibrary{shapes.geometric, arrows, positioning, fit, calc, backgrounds}
\usepackage{array}
\newcolumntype{P}[1]{>{\raggedright\arraybackslash}p{#1}}
\usepackage{booktabs} 
\tikzset{
    base/.style = {rectangle, rounded corners, draw=black, align=center, minimum height=1cm, inner sep=5pt},
    container/.style = {base, fill=blue!10, minimum width=3.5cm},
    module/.style = {base, draw=green!50!black, fill=green!10, minimum width=3cm},
    monitor/.style = {base, draw=purple!50!black, fill=purple!10, minimum width=2.5cm},
    ue/.style = {base, draw=orange, fill=orange!10, minimum width=3cm},
    arrow/.style = {thick,->,>=stealth},
    zmq/.style = {thick, <->, >=stealth, dashed, gray},
    pipe/.style     = {base, fill=blue!8,  minimum width=11.5cm, text width=11cm},
    inp/.style      = {base, fill=gray!15, minimum width=5cm},
    outbox/.style   = {base, fill=teal!15, minimum width=8cm}, 
    lbl/.style      = {font=\footnotesize\itshape, text=gray!70!black},
    parallel/.style = {base, fill=yellow!10, draw=orange!70!black,
                     minimum width=10.4cm, text width=9.2cm}
}

\graphicspath{{images/}}

\renewcommand\footnotetextcopyrightpermission[1]{}

\begin{document}

\title{SL-RFSIM: Enabling Scalable Multi-Hop 5G NR Sidelink Mesh Networking in OpenAirInterface}

\author{Simone Pio Candido}
\affiliation{
  \institution{EURECOM}
  \city{Sophia-Antipolis}
  \country{France}}
\email{Simone-Pio.Candido@eurecom.fr}

\author{Jin Yan}
\affiliation{
  \institution{EURECOM}
  \city{Sophia-Antipolis}
  \country{France}}
\email{jin.yan@eurecom.fr}

\author{Jérôme Härri}
\affiliation{
  \institution{EURECOM}
  \city{Sophia-Antipolis}
  \country{France}}
\email{jerome.haerri@eurecom.fr}

\begin{abstract}
Recent 3GPP releases have extended 5G New Radio (NR) Sidelink (SL) to support device-to-device (D2D) relay and multi-hop capabilities. This paper presents SL-RFSIM, a component-based experimentation framework extending OpenAirInterface (OAI) with scalable multi-hop NR SL capabilities. SL-RFSIM replaces the legacy OAI RF simulator with a broker-based publish/subscribe architecture enabling arbitrary peer-to-peer connectivity while preserving compatibility with the OAI protocol stack. The framework further integrates pluggable mobility, propagation, reception, and monitoring services, and supports Layer-2 mesh networking through BATMAN-adv. Experimental evaluation on the SLICES-RI research infrastructure validates the proposed architecture through representative mesh networking scenarios and identifies the current software bottlenecks limiting scalability. SL-RFSIM provides an open-source foundation for reproducible experimental research on 5G NR SL and future multi-hop cellular mesh networks.
\end{abstract}

\maketitle

\section{Introduction}
5G New Radio (NR) Sidelink (SL) has transformed cellular systems from infrastructure-centric networks into distributed communication platforms supporting direct device-to-device (D2D) connectivity. Since its initial standardization, NR SL has evolved through successive 3GPP releases to support low-latency, high-reliability applications operating beyond network coverage, while ongoing Rel.~19 activities further extend sidelink relaying and multi-hop capabilities~\cite{3gpp23287, lin2025}. Originally introduced for NR Vehicle-to-Everything (NR-V2X), 5G NR SL is now envisioned as a key enabling technology for industrial automation, collaborative robotics, public safety, unmanned systems, and tactical communications, where groups of devices exchange information directly under dynamic wireless conditions \cite{5gacia2023,qualcomm_sidelink}. In these scenarios, mesh networking and multi-hop relaying offer an effective means to extend communication range, improve resilience against infrastructure outages, and provide connectivity in sparse or obstructed environments.

Despite this growing interest, experimental evaluation of large-scale multi-hop NR SL mesh networks remains challenging. While recent experimental infrastructures (e.g. SLICES-RI~\cite{slices_blueprint} or Colosseum~\cite{bonati2021colosseum}) have considerably advanced reproducible wireless experimentation, scalable experimentation of multi-user NR SL networks integrated with executable 3GPP protocol implementations remains largely unexplored. Consequently, researchers lack an open platform for evaluating mesh networking protocols, relay mechanisms, and realistic wireless propagation under reproducible conditions.

This paper presents SL-RFSIM, a component-based experimentation framework extending OpenAirInterface (OAI)~\cite{kaltenberger2024oai} with scalable multi-hop NR SL capabilities. The contributions of this work are fourfold: (i) a broker-based RF simulation architecture supporting scalable peer-to-peer NR SL communication; (ii) an experimentation framework integrating mobility, propagation, reception and monitoring services through extensible APIs; (iii) native support for Layer-2 mesh networking through BATMAN-adv; (iv) an open-source experimental platform enabling reproducible multi-hop NR SL experimentation on local and cloud/HPC infrastructures. Experimental results validate the proposed architecture through representative multi-hop mesh scenarios, demonstrate realistic channel abstraction consistent with OAI PHY behaviour, and identify the current software bottlenecks limiting scalability.

The rest of the paper is organized as follows: Section~\ref{sec:relatedwork} reviews related work; Section~\ref{sec:OpenAirInterfaceIntro} introduces OpenAirInterface's existing sidelink and mesh architecture; Section~\ref{sec:SLRFsim} presents SL-RFSIM's modular design; Section~\ref{sec:experiment} reports experimental validation and scalability results; Section~\ref{sec:conclusion} concludes the paper.

 \vspace{-0.3cm}
\section{Related Work}
\label{sec:relatedwork}
Open-source software platforms have become essential for experimental research on cellular networks. Among them, OAI~\cite{kaltenberger2024oai} has emerged as one of the most widely adopted open-source implementations of the 3GPP 4G/5G protocol stack, supporting both software-defined radio prototyping and software-based experimentation. Recent developments have introduced support for 5G NR SL, including one-hop communication and relay functionality. Nevertheless, current implementations remain primarily intended for point-to-point communication and limited relay scenarios, providing limited support for scalable multi-user mesh experimentation.

Complementary to executable protocol stacks, numerous simulation and experimentation platforms have been developed for evaluating NR SL systems. 5G-LENA~\cite{patriciello2024lena} extends the ns-3 ecosystem with a complete 5G NR protocol model, while MoReV2X~\cite{lusvarghi2024} further introduces NR V2X sidelink capabilities for vehicular networking studies. Link-level simulators, such as {\em UWICORE NR V2X}~\footnote{\url{https://github.com/uwicore/NR-V2X-SL-LinkLevelSimulator}} provide highly accurate PHY-layer evaluation based on detailed BLER and channel models~\cite{liu2023comcom}. At a larger scale, {\em Colosseum}~\cite{bonati2021colosseum} enables hardware-in-the-loop wireless experimentation through RF emulation, whereas {\em SLICES-RI}~\cite{slices_blueprint} provides a reproducible cloud-native research infrastructure for deploying large-scale wireless experiments based on open-source software stacks. These platforms address complementary aspects of wireless experimentation, ranging from protocol simulation to reproducible infrastructure deployment, but are not specifically designed for scalable multi-hop NR SL experimentation integrated with OAI.

Mesh networking over 5G NR SL has recently attracted increasing attention from both the networking and standardization communities. Within 3GPP, Rel.~18 introduced SL relay support, while Rel.~19 continues extending relay procedures and the Sidelink Relay Adaptation Protocol (SRAP) toward more flexible distributed deployments~\cite{lin2025,3gpp38836}. At the networking layer, routing protocols such as AODV~\cite{rfc3561}, OLSR~\cite{rfc3626}, and the Linux BATMAN-adv mesh protocol~\cite{BATMANadv} have demonstrated efficient distributed route discovery for wireless mesh networks. However, these approaches generally operate independently of an executable 3GPP 5G protocol stack or are evaluated using abstract network simulators rather than real protocol implementations.

Existing solutions therefore provide complementary capabilities—including 5G NR SL models, realistic RF emulation, wireless simulation, and mesh routing—but no single framework integrates these capabilities within a composable, reproducible experimentation platform for scalable OAI-based 5G NR SL mesh networking.

 \vspace{-0.1cm}
\section{OpenAirInterface (OAI) Sidelink}\label{sec:OpenAirInterfaceIntro}
\subsection{OAI 5G NR Sidelink Architecture}
The OAI SL protocol stack supports basic 5G NR SL channels, including Physical Sidelink Broadcast Channel (PSBCH), Physical Sidelink Control Channel (PSCCH), and Physical Sidelink Shared Channel (PSSCH). SL procedures have also been included throughout the SDAP, PDCP, RLC and MAC layers in order to support 5G NR SL communication in a 2-UE scenario, either as a prototype (using USRP B210 boards) or as emulation with RFSim\footnote{\url{https://gitlab.eurecom.fr/oai/openairinterface5g/-/tree/sl-eurecom2}}. OAI SL supports mode 2 (ad-hoc) or 1-hop relay approaches.

In order to support multi-UE scenarios, OAI SL requires a scheduler, and has been extended to support a 5G NR SL mode 2(c) scheduler, while the default mode 2(a) listen-before-talk (LBT) Semi-Persistent Scheduling (SPS) is currently being implemented. Mode 2(c) corresponds to a collision-free TDMA scheduling strategy, where a coordinating UE assigns all other UEs a statically allocated set of SL subframes.  

 \vspace{-0.2cm}
\subsection{OAI Mesh Architecture}
3GPP Rel.~19 describes 5G NR Relaying architecture~\cite{3gpp38836} based on a new SRAP as well as IETF routing protocols, such as AODV~\cite{rfc3561} or OLSR~\cite{rfc3626}. While OAI 5G NR SL supports SRAP~\footnote{\url{https://gitlab.eurecom.fr/oai/openairinterface5g/-/tree/episys-sl-dev}}, APIs between SRAP and mesh protocols external to 3GPP are missing, which drove a different approach in this work. 

\begin{figure}[htbp]
\centering
\includegraphics[width=.95\columnwidth]{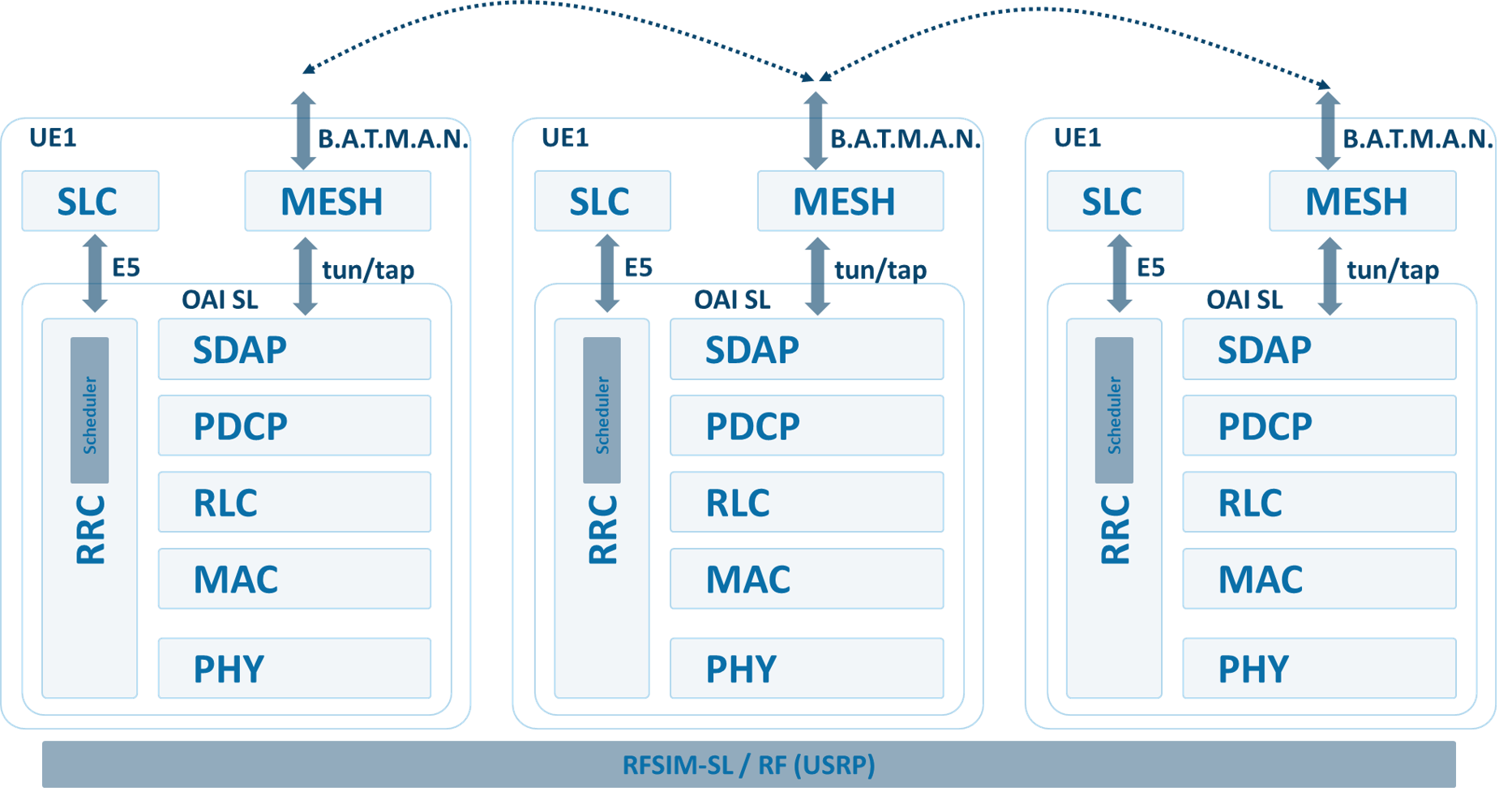}
 \vspace{-0.3cm}
\caption{OAI 5G NR SL Mesh architecture}
 \vspace{-0.4cm}
\label{fig:oai_mesh}
\end{figure}

Without loss of generality, we consider a 5G NR SL multi-hop network consisting of {\em individual} 5G NR SL wireless links, and delegate the relaying decision to higher level protocols. Rather than IETF protocols, we adopted the BATMAN-adv mesh protocol~\footnote{\url{https://www.open-mesh.org/projects/BATMAN-adv/wiki/Doc-overview}}. First, unlike IETF routing protocols, BATMAN-adv is available out of the box on any Linux kernel. Second, it operates at Layer-2, which is important for industrial wireless designs, trading SRAP fidelity for immediate availability. Fig.~\ref{fig:oai_mesh} depicts such an architecture, where the {\em Sidelink Controller (SLC)} is an entity being implemented to externally control the 5G UE RRC, notably the scheduler.   

\subsection{OAI RF Simulator (RFSim)}\label{background:rfsim}
\begin{figure}[htbp]
\centering
\includegraphics[width=.90\columnwidth]{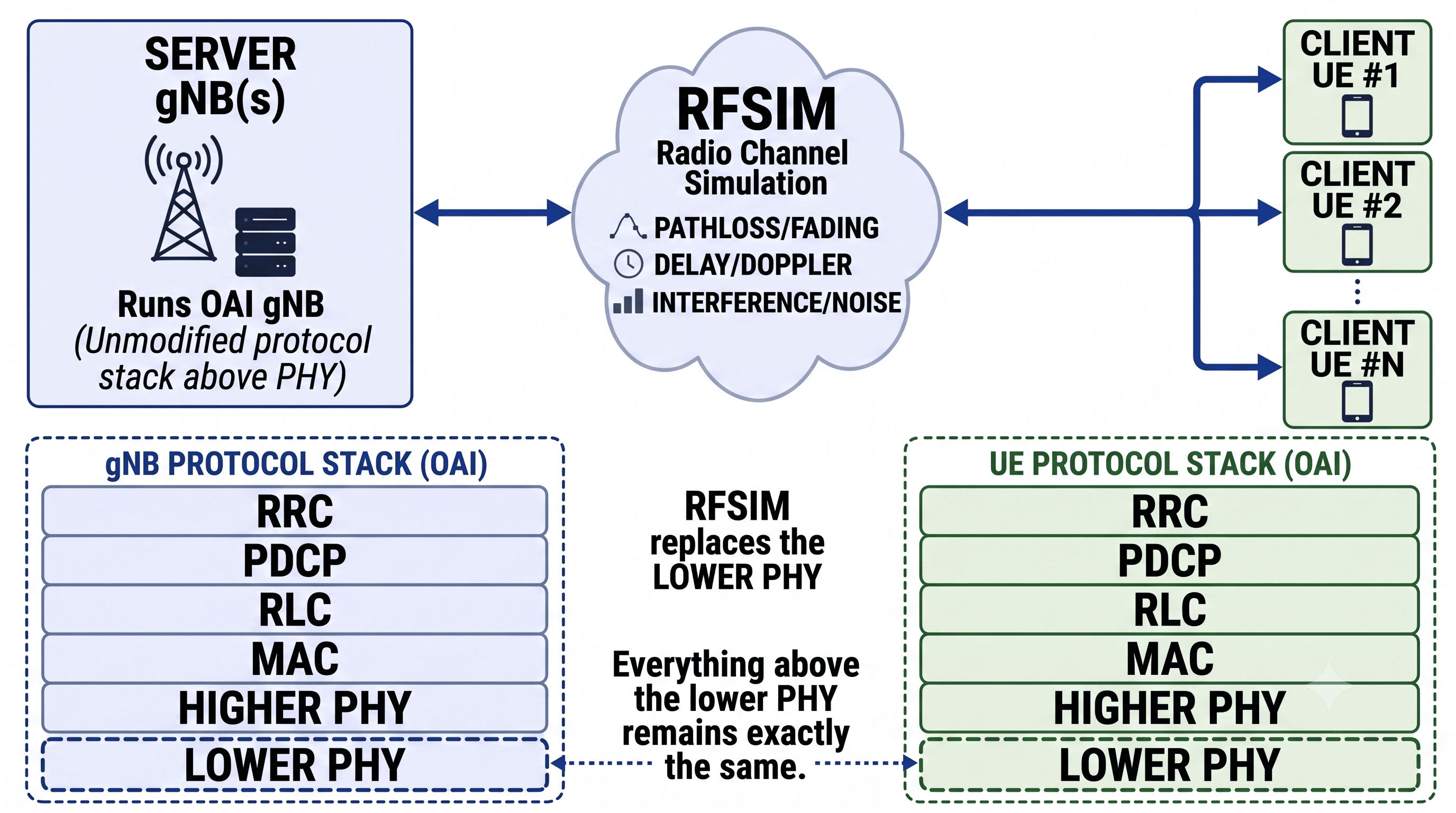}
 \vspace{-0.3cm}
\caption{Legacy OAI RFSim Architecture}
\label{fig:oai_rfsim}
\vspace{-0.5cm}
\end{figure}

The RF Simulator (RFSim) is an OAI module abstracting the lower PHY instructions and emulating the RF pipeline. Developed primarily to evaluate the impact of multi-UE connections to a gNB, RFSim follows a client-server architecture, where the server is the gNB and UEs are clients. RFSim operates as a link-level RF emulator, where channel quality or UE/gNB positions may be altered through telnet APIs, as depicted in Fig.~\ref{fig:oai_rfsim}.

Although popular for 5G NR performance evaluation, RFSim cannot be used for 5G NR SL beyond 2 UEs. In a mesh topology, RFSim clients can only be connected to one server, thus cannot directly be connected to model multi-UE SL communication.  

\begin{table*}[htbp]
\centering
\caption{Modular RF Simulator API Specification.}
\vspace{-0.3cm}
\label{tab:api_spec}
\small
\begin{tabularx}{\textwidth}{l l X }
\hline
\textbf{Module} & \textbf{Command} & \textbf{Description} \\ \hline
\textbf{Mobility} & \texttt{GET\_ALL\_POSITIONS} & Returns $x, y$ coordinates, speed, and heading for all active UEs. \\ 
(Port 5557) & \texttt{SET\_POSITION} & Manually teleports a UE, overriding internal movement logic. \\ 
& \texttt{ADD\_UE} & Dynamically registers a new UE slot in the simulation pool. \\ \hline
\textbf{Propagation} & \texttt{COMPUTE\_PATHLOSS} & Calculates attenuation ($PL_{dB}$) and Doppler shift for a TX$\to$RX link. \\ 
(Port 5558) & \texttt{SET\_MODEL} & Changes environmental logic (e.g., Urban, Highway, Rural, Indoor). \\ 
& \texttt{SET\_LINK\_BLOCK} & Forcibly blocks a directional link to simulate total obstruction. \\ \hline
\textbf{Reception} & \texttt{COMPUTE\_BER} & Maps SNR to CQI and provides a reception probability estimate. \\ 
(Port 5559) & \texttt{GET\_CQI\_TABLE} & Returns the internal 3GPP-compliant SNR-to-CQI conversion table. \\ \hline
\end{tabularx}
\end{table*}

\section{OAI SL-RFSIM}\label{sec:SLRFsim}
Considering that OAI SL experiments require a large number of UEs to model scheduling or relaying, a prototyping approach remains challenging. Accordingly, RFSim-based experiments are a perfect approach, but they need to be extended to support a meshed architecture. In this section, we describe a new RFSim tailored to 5G NR SL, which we called SL-RFSIM. 
 \vspace{-0.2cm}
\subsection{Architecture and API Specification}\label{sec:Architecture}
\begin{figure}[htbp]
 \vspace{-0.2cm}
    \centering
    \resizebox{\columnwidth}{!}{
    \begin{tikzpicture}[
        node distance=0.7cm,
        base/.style = {rectangle, rounded corners, draw=black, align=center, minimum height=0.8cm, inner sep=4pt},
        container/.style = {base, fill=blue!10, minimum width=2.6cm},
        module/.style = {base, draw=green!50!black, fill=green!10, minimum width=2.2cm},
        monitor/.style = {base, draw=purple!50!black, fill=purple!10, minimum width=1.8cm},
        ue/.style = {base, draw=orange, fill=orange!10, minimum width=2.0cm},
        arrow/.style = {thick,->,>=stealth},
        zmq/.style = {thick, <->, >=stealth, dashed, gray}
    ]
        \node (broker) [container, fill=blue!20, minimum width=3.2cm, minimum height=1.1cm] {\Large\textbf{OAI Broker} \\ \normalsize (ZMQ Proxy + Logic)};

        \node (mobility) [module, above=1.3cm of broker] {\Large\textbf{Mobility} };
        \node (propagation) [module, below left=1.2cm and 0.9cm of broker] {\Large\textbf{Propagation} };  
        \node (reception) [module, below right=1.2cm and 0.9cm of broker] {\Large\textbf{Reception}}; 

        \node (ue0) [ue, left=2.4cm of broker, line width=1.5pt] {\Large\textbf{UE 0 (Sync-Ref)}};
        \node (ue1) [ue, above=0.45cm of ue0, opacity=0.8] {\Large\textbf{ UE 1}};
        \node (uen) [ue, below=0.45cm of ue0, opacity=0.8] {\Large\textbf{ UE $n$}};

        \node (influx) [monitor, right=2.4cm of broker, yshift=0.6cm] {\Large\textbf{InfluxDB} \\ \large Time-series};
        \node (Grafana) [monitor, below=0.6cm of influx] {\Large\textbf{Grafana} \\ \large visualisation};

        \draw [arrow] (broker) -- node[anchor=west, font=\normalsize] {REQ/REP} (mobility);
        \draw [arrow] (broker) -- (propagation);
        \draw [arrow] (broker) -- (reception);

        \draw [zmq] (ue0) -- node[above, font=\normalsize, color=black] {ZMQ Pub/Sub} (broker);
        \draw [zmq] (ue1) -- (broker);
        \draw [zmq] (uen) -- (broker);

        \draw [arrow, color=purple] (broker) -- node[above, font=\normalsize] {Metrics} (influx);
        \draw [arrow, dashed, color=purple] (influx) -- (Grafana);

        \begin{scope}[on background layer]
            \node[draw, dashed, gray!60, fill=gray!5, inner sep=12pt, fit=(ue1) (uen) (mobility) (reception) (Grafana), label={[gray, font=\normalsize]above:Docker Bridge Network}] (network) {};
        \end{scope}
    \end{tikzpicture}
    }
     \vspace{-0.5cm}
    \caption{Docker architecture for OAI simulations}
     \vspace{-0.5cm}
    \label{fig:Docker_arch}
\end{figure}
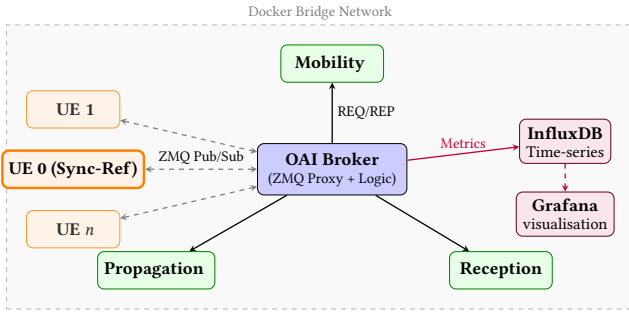

To support a full meshed topology, SL-RFSIM adopts a pub/sub approach orchestrated by a broker as depicted in Fig.~\ref{fig:Docker_arch}. It does not only connect UEs but also complementary modules required for the modelling of 5G NR experiments, such as {\em mobility}, {\em propagation}, and {\em reception} modules. An {\em InfluxDB} module, used by Grafana, has been added to provide performance monitoring of 5G NR SL experiments.  

In order to simplify runtime operation, SL-RFSIM is designed as a bridged Docker container architecture as depicted in Fig.~\ref{fig:Docker_arch}. The broker and all modules, including Grafana and InfluxDB, run on inter-connected containers. UEs may be added on-the-fly to SL-RFSIM as new containers connected to the broker.

APIs let any module be adjusted dynamically; Table~\ref{tab:api_spec} summarises the command set defined in the technical specification. UEs may be on/off-boarded on the fly in order to optimise scalability. When a UE joins, the broker automatically triggers the mobility module to initialise its trajectory, requiring no manual configuration. As per 3GPP SL specification, a {\em Sync-Ref} UE remains necessary, but it is considered as any other UE by SL-RFSIM.  

\subsection{Mobility Module}
\label{sec:mobility_module}
Mobility in 5G SL experiments is paramount. The SL-RFSIM {\em Mobility} module provides configurable mobility (position, heading and velocity) of individual UEs, updated every 100 ms. The module is based on pymobility~\footnote{\url{https://github.com/panisson/pymobility}}, a Python mobility library modelling multiple random models, ranging from Random WayPoint (RWP) and Random Walk (RWalk) to Gauss-Markov and Reference Point Group Mobility (RPGM) models. 

Without loss of generality, we apply the RWP model in this work. The parameters that define a RWP scenario are the area side length (bounding box), the admissible speed range (minimum and maximum speed in [m/s]), and the maximum pause time at each waypoint. For every leg a UE draws a speed uniformly between the minimum and maximum values. Upon reaching a waypoint it pauses for a time chosen uniformly between zero and the configured maximum.
   
The mobility module is deliberately kept separate from OAI, running as its own process or Docker container and communicating with the broker over ZMQ messaging with JSON payloads. This separation is what allows the platform to support other mobility models in the future, such as a dedicated traffic simulator like SUMO~\cite{sumo}, or a completely different domain-specific tool, without needing to redesign the broker itself. Models can also be switched at runtime, moving between Random Waypoint and Static (frozen) mode without restarting any process.

\vspace{-0.3cm}
\subsection{Propagation Module}\label{sec:propagation_module}
The SL-RFSIM propagation module is based on five 3GPP-compliant path loss models, each incorporating probabilistic LOS/NLOS selection. 
In order to handle flexible channel conditions, the module also allows setting arbitrary attenuation values for specific UE-to-UE links, or fully blocking a link, for example in the presence of walls between two UEs.

The propagation module runs as an independent process, or Docker container, separate from OAI. The broker sends it a request as JSON, and the module replies with three values: the path loss, the background noise level, and the Doppler shift for that specific transmitter-to-receiver link. The module keeps no memory between calls; at each call, 
it computes a fresh path loss considering current distance and speed, rather than tracking it over time. 

\begin{table*}[htbp]
\centering
\caption{Propagation models and path loss coefficients implemented in SL-RFSIM
         \cite{freespace, 3gpp_tr37885, 3gpp_tr38901, c2c_ecc68_2022, kryszkiewicz2024}.}
\vspace{-0.3cm}
\label{tab:prop_models}
\footnotesize
\renewcommand{\arraystretch}{1.3}
\begin{tabularx}{\textwidth}{@{}l l c >{\raggedright}X >{\raggedright\arraybackslash}X c c@{}}
\toprule
\textbf{Model} & \textbf{Standard} & \textbf{Range}
  & \textbf{PL\textsubscript{LOS} (dB)}
  & \textbf{PL\textsubscript{NLOS} (dB)}
  & \textbf{$\sigma_{SF}$}
  & \textbf{LOS/NLOS} \\
\midrule
Free Space
  & ITU-R P.525-3 & No limit
  & $20\log d + 20\log f - 27.6$
  & --
  & 0
  & LOS only \\
Urban V2V
  & 3GPP TR\,37.885 & $\le$1000\,m
  & $38.77 + 16.7\log d + 18.2\log f_c$
  & $36.85 + 30\log d + 18.9\log f_c$
  & 3\,/\,4\,dB
  & Exp. \\
Highway V2V
  & 3GPP TR\,37.885 & $\le$1000\,m
  & $32.4 + 20\log d + 20\log f_c$
  & $36.85 + 30\log d + 18.9\log f_c$
  & 3\,/\,4\,dB
  & Piecewise \\
Indoor (InH)
  & 3GPP TR\,38.901 & 1--150\,m
  & $32.4 + 17.3\log d + 20\log f_c$
  & $\max(\text{LOS},$ \allowbreak $17.3 + 38.3\log d + 24.9\log f_c)$
  & 3\,/\,8\,dB
  & Piecewise exp. \\
Rural
  & ECC-68 / Car2Car & 3-seg.
  & $d_{T0}{=}128$\,m, $n_0{=}2.8$, $d_{T1}{=}512$\,m, $n_1{=}3.3$
  & LOS only
  & --
  & LOS only \\
\bottomrule
\end{tabularx}
\end{table*}

Table~\ref{tab:prop_models} summarises the path loss coefficients, model parameters, and LOS probability types for all five models:

\textbf{Free Space} follows ITU-R P.525-3 \cite{freespace} and serves as a theoretical baseline with no LOS/NLOS selection; it is used to isolate the effect of distance without environmental effects.

\textbf{Urban V2V} (3GPP TR 37.885 \cite{3gpp_tr37885}) defines three channel states - LOS, NLOSv (vehicle blocker), and NLOS (building blocker). Each state is selected probabilistically per query according to a density-dependent exponential function \cite{ns3_v2v_prob_src}; the  $A_{\text{NLOSv}}$ depends on the relative heights of the transmitter, receiver, and blocking vehicle, and is calculated using one of three cases~\cite{ns3_v2v_pl_src}.

\textbf{Highway V2V} (3GPP TR 37.885 \cite{3gpp_tr37885}) uses the same path loss expressions as Urban, but the probability of LOS follows a different curve: it decays gradually beyond 475\,m~\cite{ns3_v2v_prob_src}; the same $A_{\text{NLOSv}}$ model from Urban V2V applies here too~\cite{ns3_v2v_pl_src, morev2x}.

\textbf{Indoor (InH)} (3GPP TR 38.901 \cite{3gpp_tr38901}) applies a piecewise exponential LOS probability valid from 1 to 150\,m and takes $PL_{\text{NLOS}} = \max(PL_{\text{LOS}},\, PL'_{\text{NLOS}})$ to enforce the physical constraint that NLOS loss cannot be lower than LOS \cite{poddar2025}.

\textbf{Rural} adopts the ECC Report 68 three-segment formula as parametrised by the Car2Car Consortium for vehicle-height antennas~\cite{c2c_ecc68_2022, kryszkiewicz2024}, and independently validated~\cite{garciavidal2025mco}.

 \vspace{-0.2cm}
\subsection{Reception Module}
\label{sec:reception_module}
Similar to RFSim, SL-RFSIM only abstracts the lower PHY instructions. Accordingly, OAI UEs contain functions to compute packet reception from IQ samples provided by SL-RFSIM. However, the OAI UEs do not contain hooks sending channel quality metrics or reception conditions to external entities. The actual packet reception is decided by OAI's internal PHY layer on real IQ samples, while the external reception module provides a parallel, configurable estimate, usable today for observability, and designed to be extended toward driving system-level reception decisions.

SL-RFSIM reception module measures the channel quality metrics using a specifically designed pipeline, and computes the Block Error Rate (BLER) from Signal-to-Noise Ratio (SNR) according to the UWICORE NR~V2X~SL dataset, which contains 6{,}188 BLER vs.\ SNR curves following 3GPP Rel.~16 specifications~\cite{lusvarghi2024}.

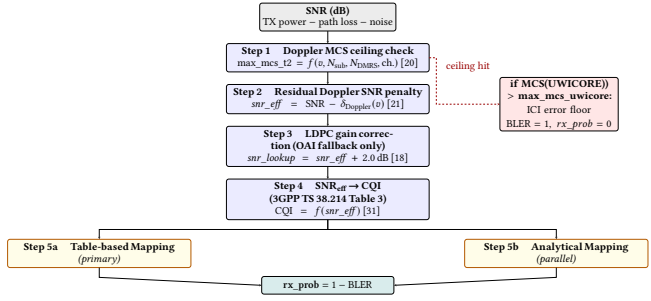
\begin{figure}[t]
\centering
\resizebox{\columnwidth}{!}{%
\begin{tikzpicture}[node distance=0.7cm and 1.0cm]

  \node[inp] (snr)
    {\Huge\textbf{SNR (dB)}\\[2pt]
     \Huge TX power $-$ path loss $-$ noise};

  \node[pipe, below=of snr] (s1)
    {\Huge\textbf{Step 1 \quad Doppler MCS ceiling check}\\[2pt]
     \Huge $\text{max\_mcs\_t2} = f(v,N_\text{sub},N_\text{DMRS},\text{ch.})$ \cite{lusvarghi2024vtc}};

  \node[draw, fill=red!10, rounded corners, text width=6.5cm,
        align=center, below right=0.3cm and 4cm of s1, font=\Huge] (ceil)
    {\Huge\textbf{if MCS(UWICORE)) $>$ max\_mcs\_uwicore:} \\[2pt]
     ICI error floor \\[2pt] 
     $\mathrm{BLER}=1,\ rx\_prob=0$}; 

  \node[pipe, below=of s1] (s2)
    {\Huge\textbf{Step 2 \quad Residual Doppler SNR penalty}\\[2pt]
     \Huge $\mathit{snr\_eff} = \mathrm{SNR} - \delta_{\mathrm{Doppler}}(v)$ \cite{lusvarghi2024}};

  \node[pipe, below=of s2] (s3)
    {\Huge\textbf{Step 3 \quad LDPC gain correction (OAI fallback only)}\\[2pt]
     \Huge $\mathit{snr\_lookup} = \mathit{snr\_eff} + 2.0\,\mathrm{dB}$ \cite{liu2023comcom}};

  \node[pipe, below=of s3, text width=8cm] (s4)
    {\Huge\textbf{Step 4 \quad SNR$_\text{eff}$ $\boldsymbol{\to}$ CQI \quad (3GPP TS\,38.214 Table\,3)}\\[3pt]
     \Huge $\mathrm{CQI} = f(\mathit{snr\_eff})$ \cite{serunin2022}};

   \node[parallel, below left=1.1cm and 2.0cm of s4] (s5a)
    {\Huge\textbf{Step 5a \quad Table-based Mapping}\\[2pt]
     \textit{(primary)}};

  \node[parallel, below right=1.1cm and 2.0cm of s4] (s5b)
    {\Huge\textbf{Step 5b \quad Analytical Mapping}\\[2pt]
     \textit{(parallel)}};

  \node[outbox, minimum width=7.5cm,
        at={($(s5a.south)!0.5!(s5b.south)+(0,-1.0cm)$)}] (rxp)
    {\Huge $\mathbf{rx\_prob} = 1 - \mathrm{BLER}$};

  \draw[arrow] (snr) -- (s1);
  \draw[arrow] (s1)  -- (s2);
  \draw[arrow] (s2)  -- (s3);
  \draw[arrow] (s3)  -- (s4);
  \draw[arrow] (s4.south) -- ++(0,-0.55) -| (s5a.north);
  \draw[arrow] (s4.south) -- ++(0,-0.55) -| (s5b.north);
  \draw[arrow] (s5a.south) -- ($(s5a.south)+(0,-0.55)$) -- (rxp.west);
  \draw[arrow] (s5b.south) -- ($(s5b.south)+(0,-0.55)$) -- (rxp.east);

  \draw[arrow, dashed, red!60!black]
    (s1.east) -- ++(0.4,0) |- (ceil.west);
  \node[font=\Huge, red!60!black] at ($(s1.east)+(2.2,-0.5)$) {ceiling hit};

\end{tikzpicture}%
}
\vspace{-0.5cm}
\caption{Reception Probability Pipeline}
\vspace{-0.5cm}
\label{fig:rx_pipeline}
\end{figure}

The full pipeline converting SNR to a final reception probability, $rx\_prob$, works through six steps: a Doppler ceiling check, an SNR penalty, CQI mapping, MCS mapping, a BLER lookup, and finally the reception probability itself. This procedure is illustrated in Fig.~\ref{fig:rx_pipeline}. 

\vspace{-0.2cm}
\paragraph{Doppler MCS Ceiling}\label{sec:doppler_ceiling}
Beyond a certain speed, CP-OFDM's inter-carrier interference (ICI) makes the link fail regardless of signal strength. We adopt this~\textit{ceiling} from Lusvarghi et al.~\cite{lusvarghi2024vtc}, the highest MCS reliable for a given channel and speed . Exceeding it sets $\text{BLER}=1$, $\text{rx\_prob}=0$ directly (Fig.~\ref{fig:rx_pipeline}, Step~1).

\vspace{-0.2cm}
\paragraph{Residual Doppler SNR Penalty}\label{sec:doppler_penalty}
Below the ceiling, Doppler still degrades performance as added noise, modelled as a speed-dependent penalty applied before all subsequent steps~\cite{lusvarghi2024} (Fig.~\ref{fig:rx_pipeline}, Step~2).

\vspace{-0.2cm}
\paragraph{LDPC Gain Correction}\label{sec:LDPC_gain} 
UWICORE already reflects the LDPC coding gain, having been measured from real coded transmissions; OAI's fallback tables do not. We apply a fixed 2.0\,dB correction only when using the fallback tables, skipping it for UWICORE where the gain is already built in (Fig.~\ref{fig:rx_pipeline}, Step~3)~\cite{liu2023comcom}. 

\vspace{-0.2cm}
\paragraph{SNR to CQI}\label{sec:snr2cqi}
The effective SNR is mapped to CQI following the same 3GPP TS~38.214 table OAI uses for sidelink scheduling~\cite{3gpp_ts38214}, listed in full in Table~\ref{tab:snr_cqi_mcs}.

\vspace{-0.2cm}
\paragraph{CQI to MCS} \label{sec:cqitomcs}
OAI and UWICORE index MCS differently for sidelink; this step just matches entries between the two tables so the correct MCS reaches the correct BLER dataset (Table~\ref{tab:snr_cqi_mcs}). 

\vspace{-0.2cm}
\paragraph{BLER and Reception Probability}
Given channel type, speed, and MCS, we interpolate the matching BLER value in the UWICORE dataset at the effective SNR, then compute $P_{rx} = 1 - \mathrm{BLER}$.

\begin{table}[htbp]
\vspace{-0.2cm}
\centering
\caption{Combined SNR-to-CQI and CQI-to-MCS mapping }
\vspace{-0.3cm}

\label{tab:snr_cqi_mcs}
\footnotesize
\renewcommand{\arraystretch}{1.15}
\resizebox{\columnwidth}{!}{%
\begin{tabular}{@{}cclcccc@{}}
\toprule
\textbf{CQI} & \textbf{SNR (dB)} & \textbf{Mod.} & \textbf{CR/1024} & \textbf{Eff.} & \textbf{MCS(OAI)} & \textbf{MCS(UWICORE)} \\ \midrule
0   & $< -20$      & -   & -  & -    & - & -  \\
1   & $[-20,-6)$   & QPSK  &  30  & 0.0586 & - & -  \\
2   & $[-6,-3)$    & QPSK  &  50  & 0.0977 & - & -  \\
3   & $[-3,0]$     & QPSK  &  78  & 0.1523 & - & -  \\
4   & $(0,2]$      & QPSK  & 120  & 0.2344 &  6  &  0   \\
5   & $(2,3]$      & QPSK  & 193  & 0.3770 &  8  &  1   \\
6   & $(3,5]$      & QPSK  & 308  & 0.6016 & 10  &  2   \\
7   & $(5,7]$      & QPSK  & 449  & 0.8770 & 12  &  3   \\
8   & $(7,9]$      & QPSK  & 602  & 1.1758 & 14  &  4   \\
9   & $(9,10]$     & 16QAM & 378  & 1.4766 & 16  &  5   \\
10  & $(10,12]$    & 16QAM & 490  & 1.9141 & 18  &  7   \\
11  & $(12,15]$    & 16QAM & 616  & 2.4063 & 20  &  9   \\
12  & $(15,16]$    & 64QAM & 466  & 2.7305 & 22  & 11   \\
13  & $(16,18]$    & 64QAM & 567  & 3.3223 & 24  & 13   \\
14  & $(18,19]$    & 64QAM & 666  & 3.9023 & 26  & 15   \\
15  & $> 19$       & 64QAM & 772  & 4.5234 & 28  & 17   \\ \bottomrule
\end{tabular}%
}
\vspace{-0.5cm}
\end{table}

\subsection{Broker}\label{sec:Broker}
The broker is the main orchestrator of SL-RFSIM: simple, lightweight, aware of which UEs are connected and of the channel conditions between each pair. Its most crucial tasks are (i) \emph{receiving subscriptions and publishing on the subscribed UE channels}, (ii) {\em orchestrating data flow} between the SL-RFSIM modules on the forwarded IQ samples: path loss scaling, Doppler rotation, and SNR estimation (Sections~\ref{sec:mcl}, \ref{sec:doppler_iq}, \ref{sec:analytical_snr}), computed independently per UE pair and published on its own dedicated channel.

Fig.~\ref{fig:seq_all} illustrates this workflow. During sync/UE join and direct UE-to-UE mesh links (Fig.~\ref{fig:seq_join_mesh}), the broker caches UE positions every 100 ms, queries the propagation module for path loss and Doppler shift, applies Doppler phase rotation and MCL-based path loss scaling to the IQ samples, then forwards them to the receiver. It also periodically queries the reception module for CQI, BER, and reception estimates, used for monitoring (Fig.~\ref{fig:seq_all}b). Actual reception, however, remains entirely within OAI's PHY layer, never duplicated by the broker.

\subsubsection{Differential Path-loss Scaling and I/Q Quantization}\label{sec:mcl}
This section details the path loss scaling step from Fig.~\ref{fig:seq_join_mesh}: how the broker applies the propagation path loss value to the actual IQ samples. OAI represents each I/Q sample as a 16-bit integer with amplitude $\approx 256$ which can cause numerical truncation to zero even at short range. 

To avoid this, the broker adopts the Minimum Coupling Loss (MCL) convention \cite{3gpp_tr36942}, exposed as a runtime CLI parameter (\texttt{--mcl}):
\[
  PL_{\text{effective}} = \max(0,\, PL_{\text{actual}} - \text{MCL})
\]

This effective attenuation in decibels is then converted into a linear scalar factor and applied to the I/Q samples:
$$s_{out}[n] = s_{in}[n] \times 10^{\frac{-PL_{\text{effective}}}{20}}$$

Empirical calibration during this project demonstrated that setting the MCL to $80\,\text{dB}$ optimally aligns the I/Q sample amplitude with the theoretical SNR, minimising the divergence between the link-level OAI decoder and the system-level UWICORE tables.

\subsubsection{OFDM-Aware Doppler Phase Rotation}\label{sec:doppler_iq}
This section details the Doppler rotation step from Fig.~\ref{fig:seq_join_mesh}. UE movement causes Doppler shift as mentioned before in Section~\ref{sec:doppler_ceiling}. We model this by rotating the phase forward slightly at each OFDM symbol, then resetting it to zero at the start of every 2 ms block, consistent with OAI RFSim's assumption of no residual carrier frequency offset (CFO) carried over between blocks on the sidelink path: 
\begin{align*}
  \varphi_k &= \varphi_{k-1} + 2\pi f_d \cdot T_{\text{sym},k} \\
  s_\text{out}[n] &= s_\text{in}[n] \cdot e^{j\varphi_k}
    \quad \forall n \in \text{symbol } k
\end{align*}

Tested on a highway scenario with two UEs positioned 200 m apart, PSBCH success drops from $\approx$100\% with no channel effects to $\approx$73\% with Doppler off, recovering to $\approx$81\% at $f_d \approx 1$\,Hz; PSCCH/PSSCH decoding remains largely unaffected thanks to DM-RS-based correction.
We reset the phase every block rather than letting it accumulate, since OAI's sidelink RFSim has no pilot-based CFO correction on the receive path; without a reset, drift builds up faster than the receiver can track, breaking synchronisation within a few hundred milliseconds. 

\subsubsection{Analytical SNR Estimation}\label{sec:analytical_snr}
\begin{figure}[htbp]
    \centering
    \begin{subfigure}[t]{0.34\textwidth}
        \centering
        \includegraphics[width=\textwidth]{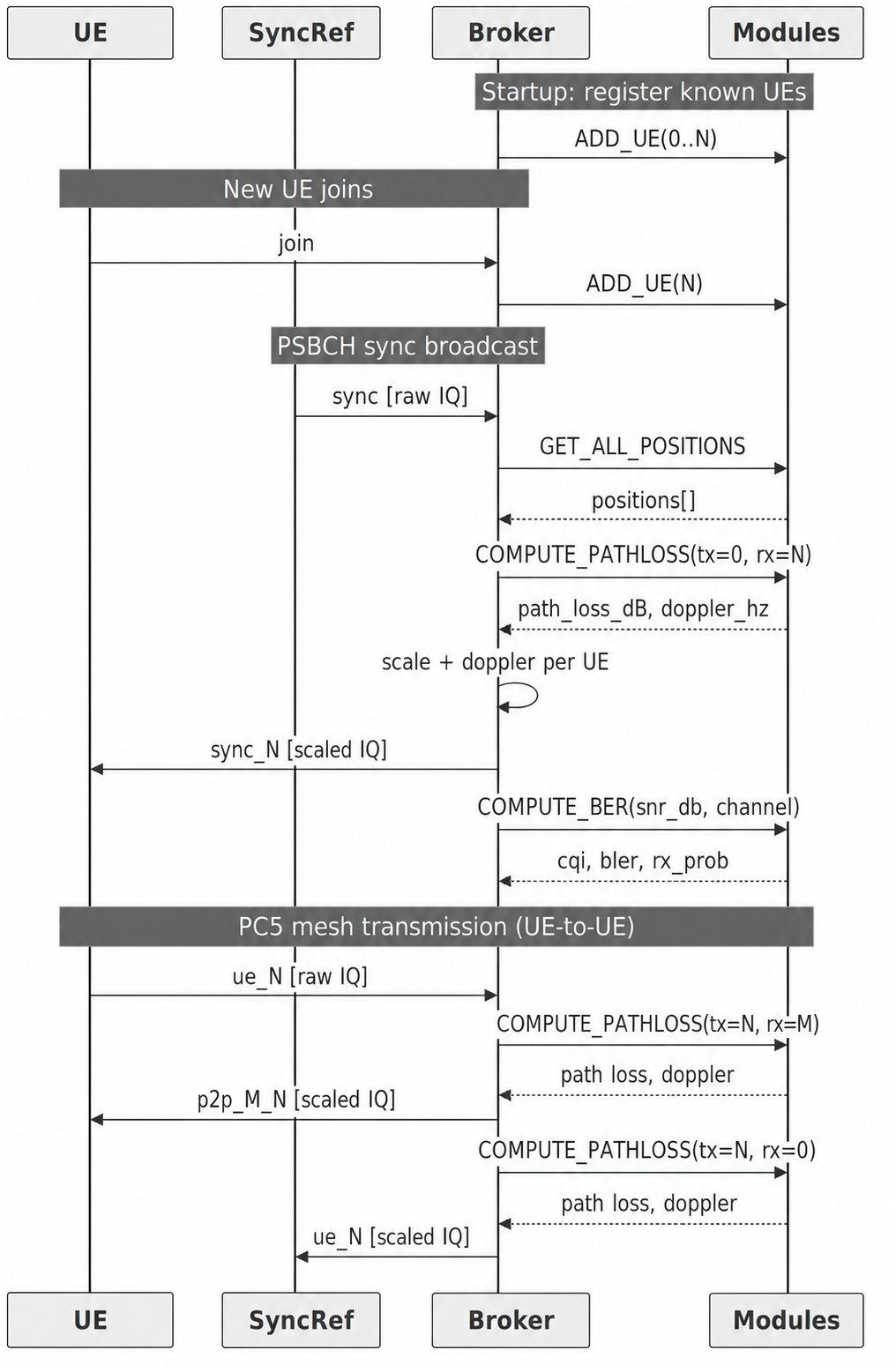}
        \caption{UE join, sync broadcast, and PC5 mesh transmission.}
        \label{fig:seq_join_mesh}
    \end{subfigure}
    \hfill
    \begin{subfigure}[t]{0.34\textwidth}
        \centering
        \includegraphics[width=\textwidth]{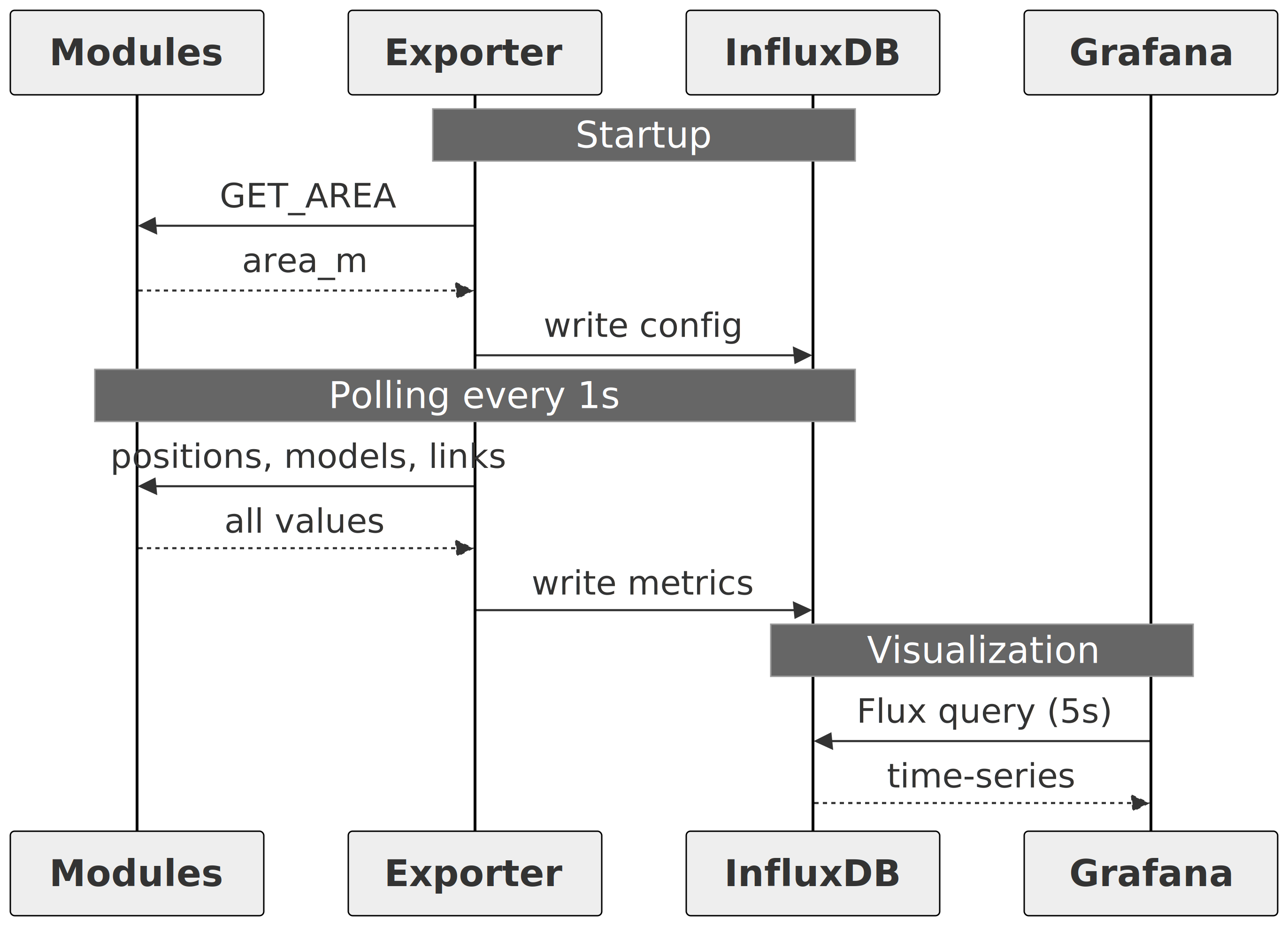}
        \caption{Dashboard monitoring.}
        \label{fig:seq_dashboard}
    \end{subfigure}
    \vspace{-0.4cm}
    \caption{Sequence diagrams of the broker's main communication patterns.}
    \label{fig:seq_all}
    \vspace{-0.4cm}
\end{figure}

This section covers the final step from Fig.~\ref{fig:seq_join_mesh}: the broker's SNR calculation, derived from path loss rather than measured directly:  
$$SNR_{dB} = P_{TX} - PL_{dB} - (N_0 + 10 \log_{10}(BW))$$

where $P_{\mathrm{TX}} = 23\,\mathrm{dBm}$, $N_0 = -174\,\mathrm{dBm/Hz}$, and $\mathrm{BW}$ is the operational bandwidth.~\footnote{This deliberately uses the unmodified path loss, not the MCL-clipped value from Section~\ref{sec:mcl}; applying it here would inflate SNR and invalidate the reception estimates in Sections~\ref{sec:reception_module} and~\ref{sec:cqires}.}

 \vspace{-0.2cm}
\subsection{Grafana \& InfluxDB Monitoring Stack}
Grafana~\cite{Grafana_docs} is a popular visualisation layer for its native support for time-series data sources and real-time dashboard rendering, backed by InfluxDB \cite{ionos_influxdb} as the time-series database for its high write throughput and native support for timestamped metrics at sub-second granularity. We therefore integrated Grafana as SL-RFSIM's visualisation layer.
Figure~\ref{fig:seq_dashboard} shows how this stack is populated. An Exporter polls each module every second and writes to InfluxDB, which Grafana reads via 5 s Flux queries. This polling-based design keeps the monitoring stack fully decoupled from OAI itself; Grafana never queries OAI or the broker directly, only InfluxDB.
The SL-RFSIM broker therefore periodically publishes monitoring KPIs to InfluxDB, which the SL-RFSIM Grafana dashboard may display (Details in Section~\ref{sec:Grafana_demo}).

 \vspace{-0.2cm}
\section{Experimentation \& Validation}\label{sec:experiment}
The experimental validation of SL-RFSIM 
has been conducted on the SLICES-RI~\footnote{\url{https://www.slices-ri.eu/}} OpenXGLab node at EURECOM, using an x86\_64 cluster with 8 Xeon Gold CPU cores and 8 GB of RAM.   

\subsection{Cluster Access \& Scenario Configuration}
The OAI 5G NR SL mesh experiment with SL-RFSIM is performed through a Jupyter notebook environment, which is accessed with a SLICES-RI account~\footnote{\url{https://jupyterhub.slices.cs.eurecom.fr/}}. Source code is cloned from a specific {\em module3-batmanmoreues} branch from the {\em GitLab}~\footnote{\url{https://gitlab.eurecom.fr/oai-sl-mobility/oai-sl-mobility}}. In order to avoid recompiling OAI source code for each container UE, all containers are provided ready-to-use through {\em Docker Hub} via preconfigured scripts available in the OAI SL-RFSIM folder~\footnote{\url{radio/rfsim++/examples}}.

Scenarios consist of Python scripts pre-configuring all SL-RFSIM aspects (mobility, propagation, number of UEs, and active mesh links) via the API in Table~\ref{tab:api_spec}, as well as scripts to start all containers required by the experiment. This validation focuses on functional correctness and scalability limits of the broker architecture; broker-latency~\footnote{The reception module's latency does not impact SL-RFSIM OAI UE reception.} characterisation and a direct baseline comparison against legacy client-server RFSim are left to future work.

 \vspace{-0.2cm}
\subsection{Grafana Dashboard}\label{sec:Grafana_demo}
\begin{figure}[htbp]
\centering
\begin{subfigure}[t]{1\columnwidth}
    \centering
    \includegraphics[width=\textwidth]{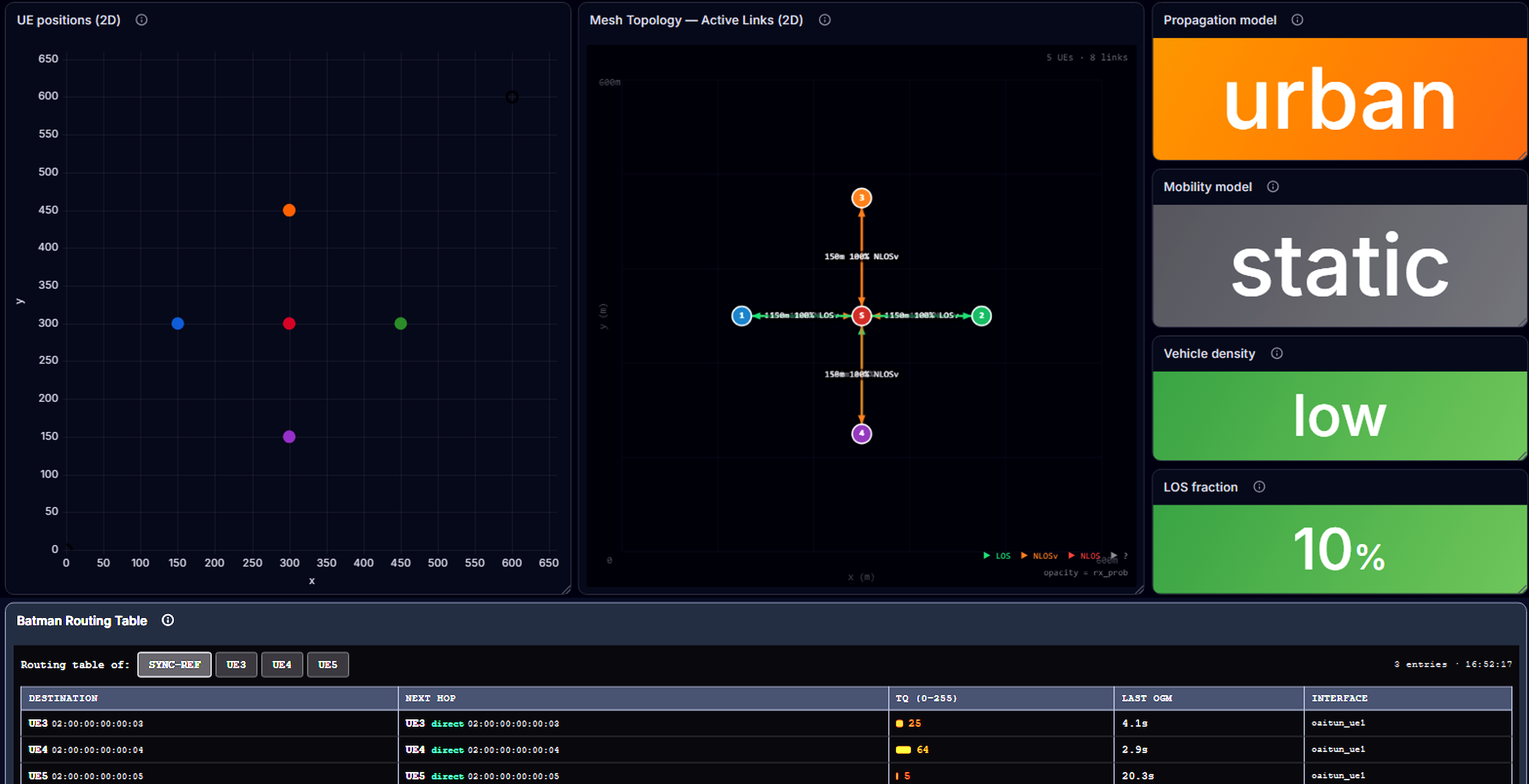}
    \caption{UE positions, mesh topology and Batman Routing Status}
    \label{fig:graf_upper}
\end{subfigure}
\hfill
\begin{subfigure}[t]{1\columnwidth}
    \centering
    \includegraphics[width=\textwidth]{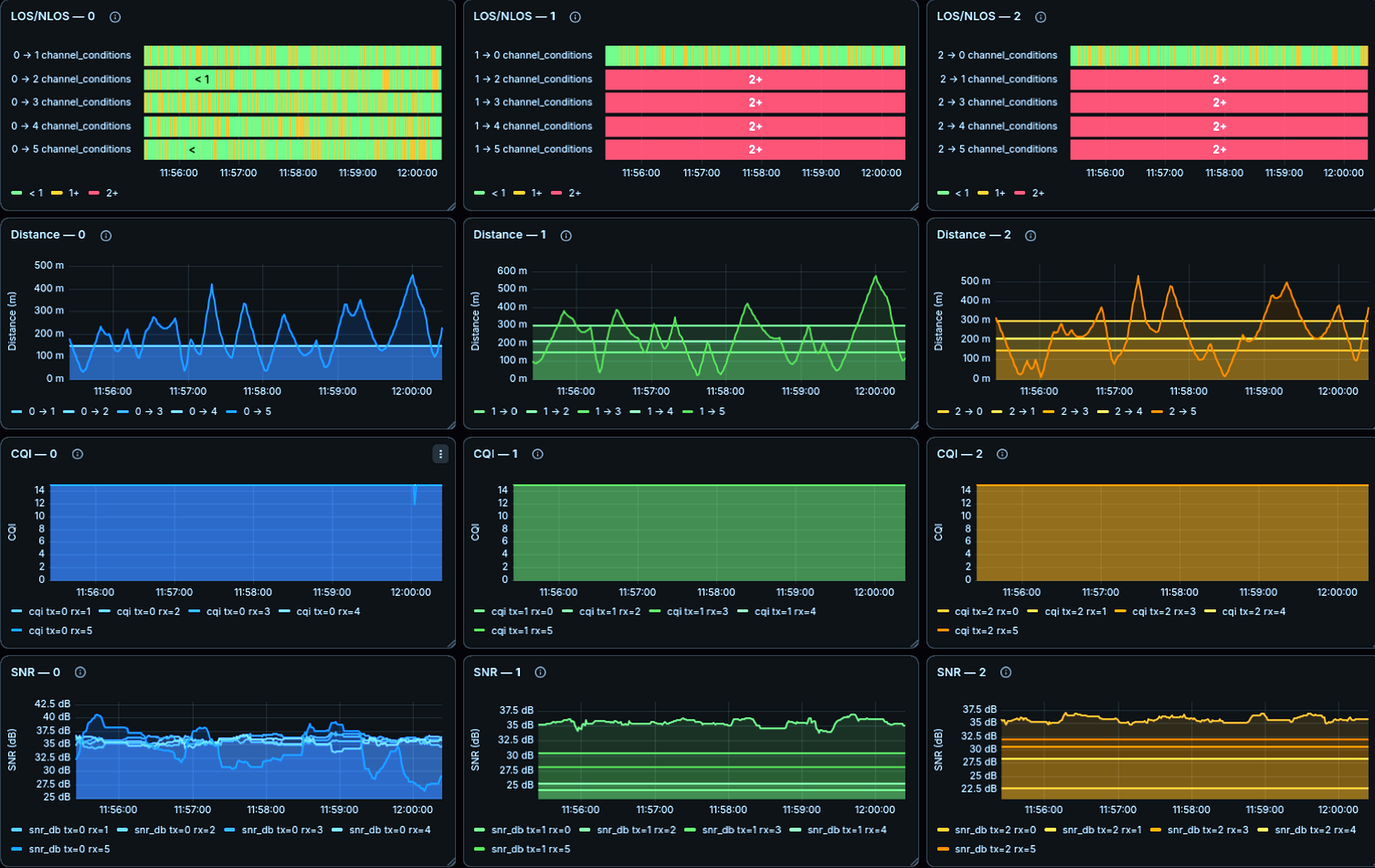}
    \caption{per-UE LOS/NLOS state, distance, CQI and SNR time-series}
    \label{fig:graf_middle}
\end{subfigure}
\hfill
\begin{subfigure}[t]{1\columnwidth}
    \centering
    \includegraphics[width=\textwidth]{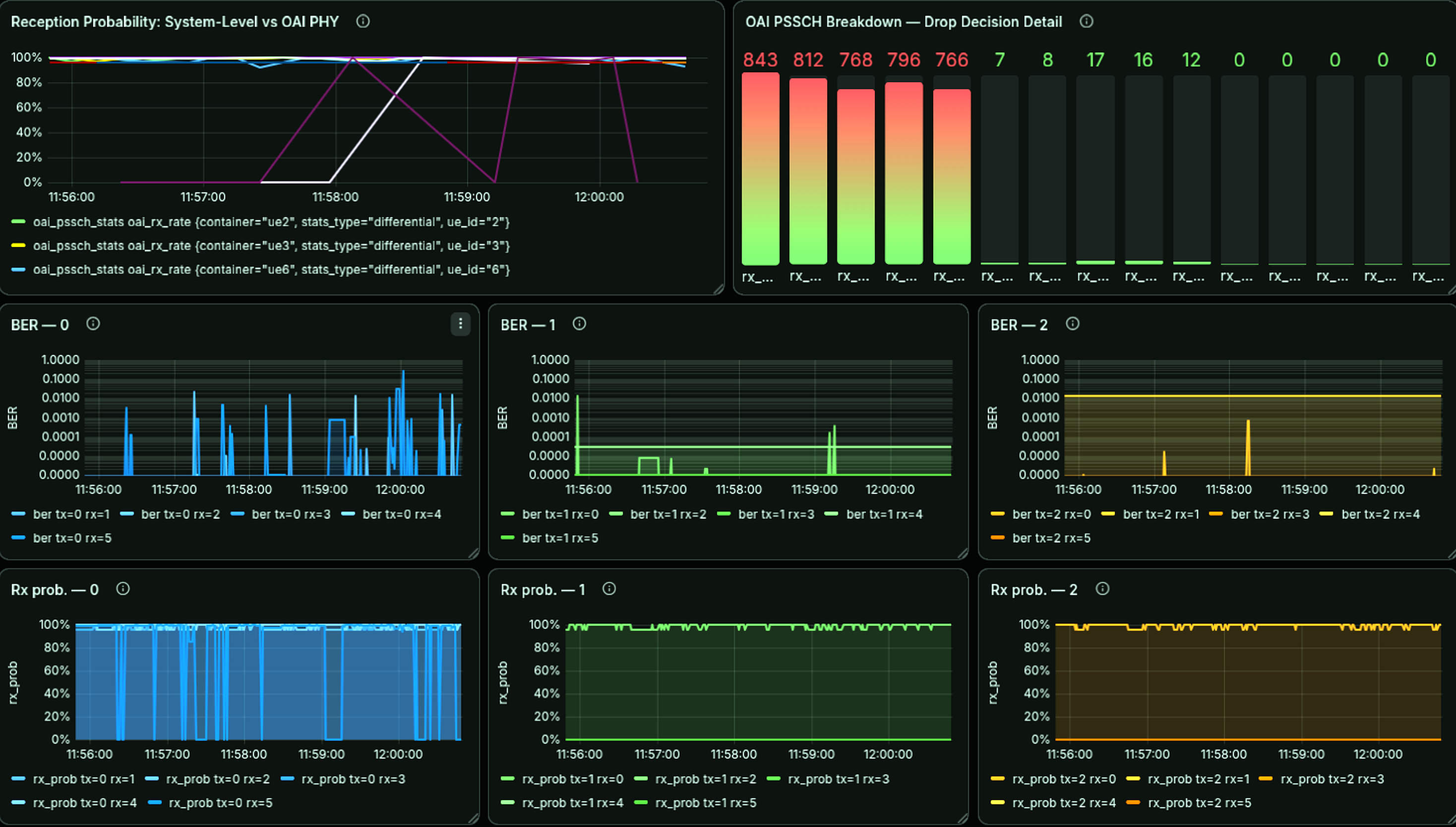}
    \caption{system-level vs. PHY-layer reception probability and BER}
    \label{fig:graf_bottom}
\end{subfigure}
\vspace{-0.5cm}
\caption{SL-RFSIM Grafana Dashboard}
\label{fig:Grafana}
\vspace{-0.5cm}
\end{figure}

The dashboard is organised around link pairs (UEi\,$\to$\,UEj); a dropdown selects the reference UE for all four time-series panels: (i) a 2D map with real-time UE positions, directional links, and LOS/NLOS state per pair; (ii) per-link SNR, CQI, BER, and BLER time-series with UE selection via dropdown; (iii) PSCCH statistics comparing the layer-2 reception lock ratio against the system-level model prediction; (iv) distance and LOS fraction panels for correlation between physical proximity and channel metrics. The topology view (Fig.~\ref{fig:Grafana}) shows real-time UE positions and active links; per-link metrics (SNR, CQI, BER, $rx\_prob$) are shown in Figs.~\ref{fig:graf_middle} and~\ref{fig:graf_bottom}.

\vspace{-0.2cm}
\subsection{Reception Module Validation}
\label{sec:cqires}
In SL-RFSIM, the reception module computes CQI externally from the broker's analytical SNR for monitoring purposes, while OAI UEs compute CQI internally at the upper PHY layer. A divergence is therefore possible: SL-RFSIM reception module may report high $\text{rx\_prob}$ while OAI's polar decoder would drop the packet. Table~\ref{tab:rx_comparison} quantifies this gap for two highway scenarios with MCL\,=\,200\,dB (IQ passthrough) and shadow fading enabled ($\sigma_{\mathrm{SF}}=3$\,dB, TR\,37.885). 

\begin{table}[htbp]
\centering
\small
\caption{System-level vs.\ link-level reception probability.}
\vspace{-0.3cm}
\label{tab:rx_comparison}
\begin{tabular}{lcccc}
\hline
Scenario & $p_{\mathrm{rx}}^{\mathrm{sys}}$ & $\hat{p}_{\mathrm{rx}}^{\mathrm{OAI}}$ & $|\Delta|$ & Agreement \\
\hline
1000\,m highway & 97.59\% & 96.64\% & 0.95\,pp & 99.1\% \\
5000\,m highway & 91.98\% & 93.27\% & 1.29\,pp & 98.7\% \\
\hline
\end{tabular}
\end{table}

The gap turns out to be small enough to be negligible for monitoring purposes. At $1000\,\mathrm{m}$, an NLOSv-dominated scenario (CQI\,11--15), agreement is 99.1\%; at $5000\,\mathrm{m}$, closer to the sensitivity threshold (CQI\,4--7), agreement remains as high as 98.7\%, despite high channel variability ($\sigma_{p_{\mathrm{rx}}} = 14.9\%$).

 \vspace{-0.2cm}
\subsection{Multi-hop Line Topology}
\begin{figure}[htbp]
    \centering
    \begin{subfigure}[t]{0.28\columnwidth}
        \centering
        \includegraphics[width=\linewidth]{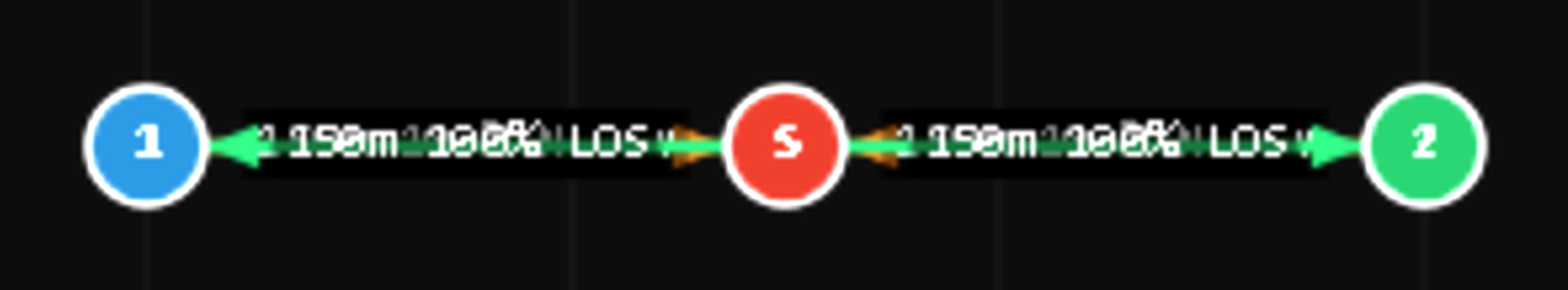}
        \caption{Line Topology}
        \label{fig:multihop_topology}
    \end{subfigure}
    \hfill
    \begin{subfigure}[t]{0.28\columnwidth}
        \centering
        \includegraphics[width=\linewidth]{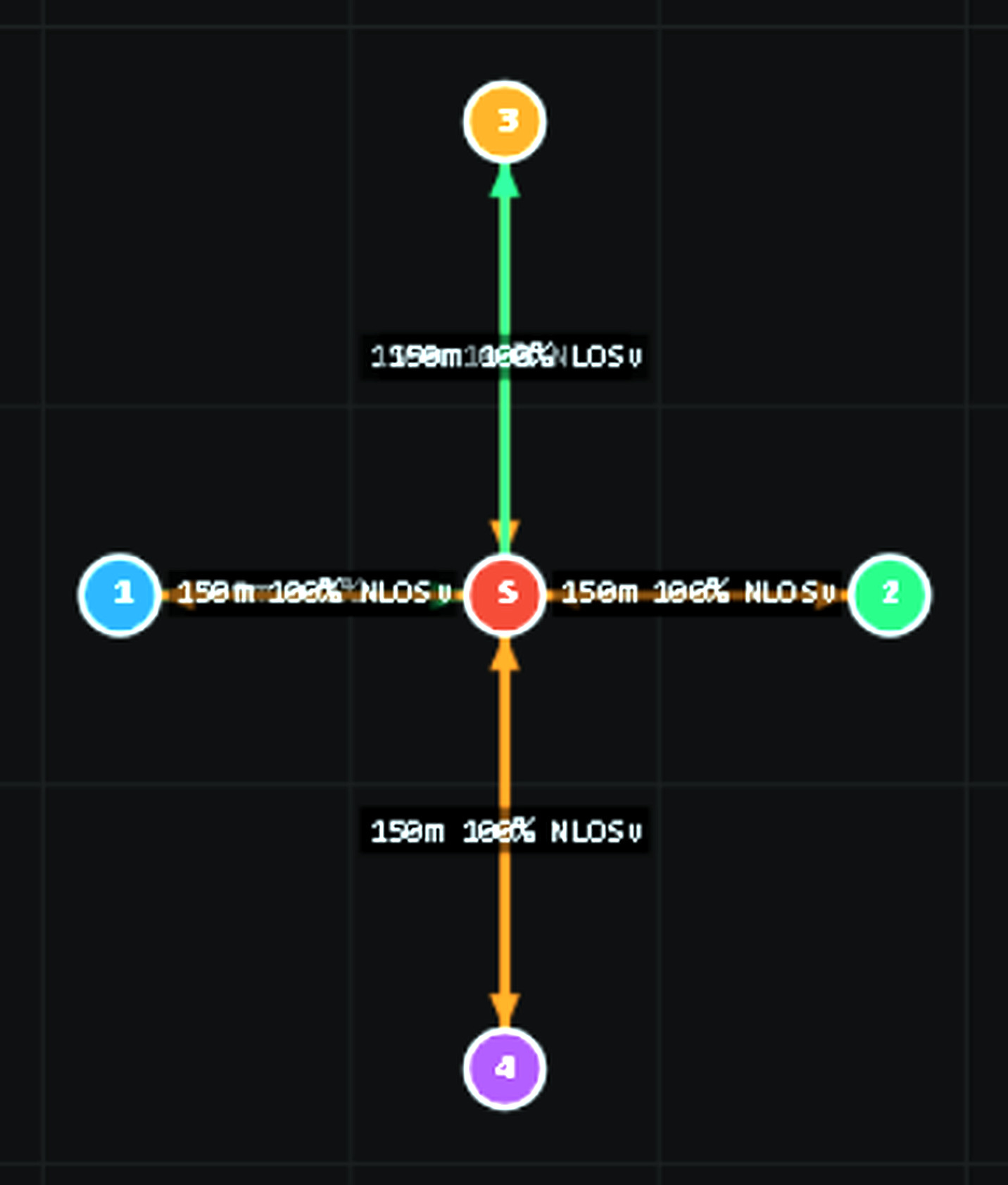}
        \caption{Star Topology}
        \label{fig:star_topology_map}
    \end{subfigure}
    \hfill
    \begin{subfigure}[t]{0.3\columnwidth}
        \centering
        \includegraphics[width=\linewidth]{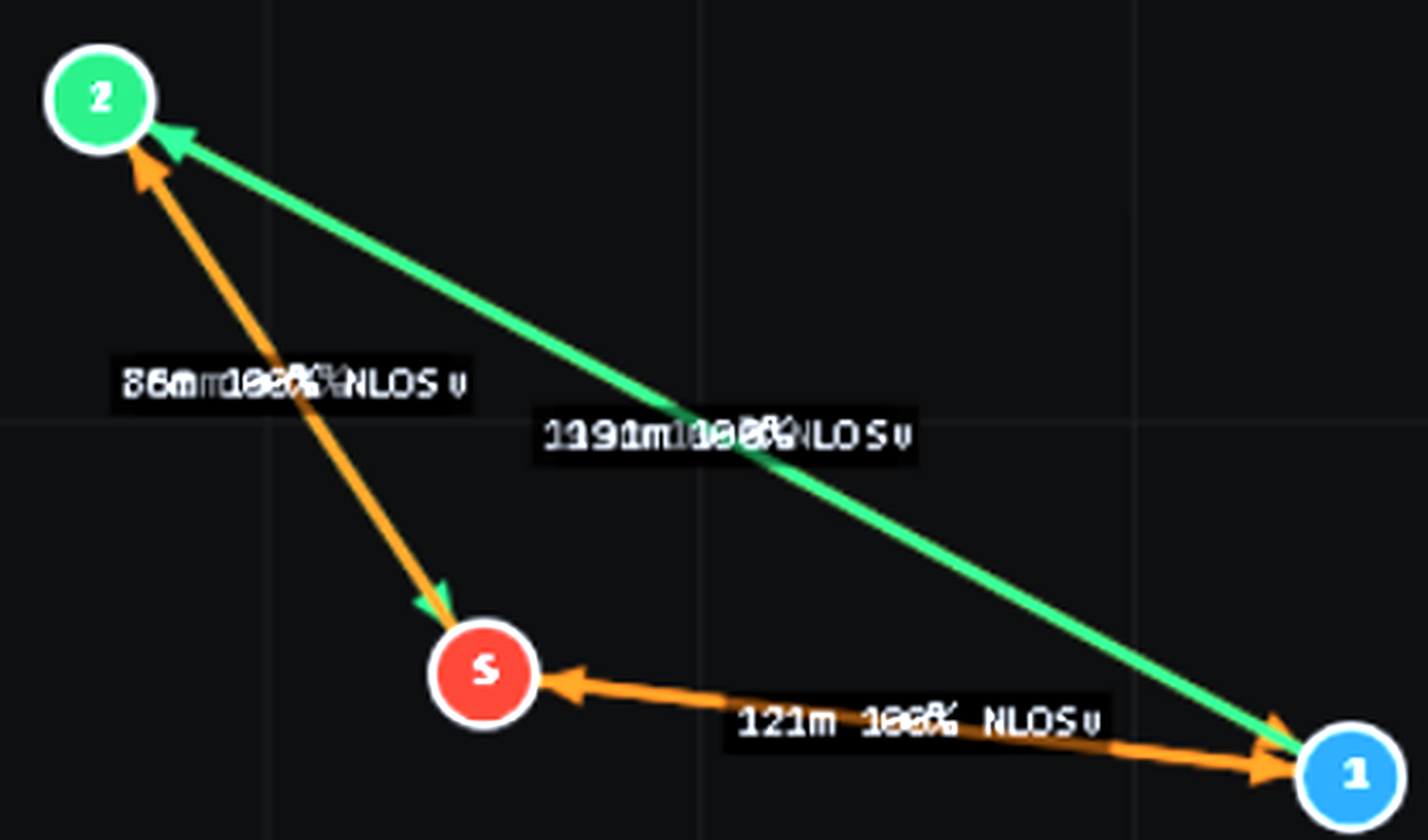}
        \caption{Mobile Topology}
        \label{fig:star_mobile_topology}
    \end{subfigure}
    \vspace{-0.3cm}
    \caption{Grafana topology panels for the three scenarios.}
    \label{fig:topology_all}
    \vspace{-0.5cm}
\end{figure}

The primary objective of this scenario is to verify multi-hop relaying: that traffic between two UEs correctly routes through an intermediate node with no direct link between them. Three UEs were arranged in a line, UE0 (the Sync-Ref) at the centre, with UE1 and UE2 placed 150 m away on either side. Per-link channel conditions are shown in Table~\ref{tab:link_quality}. The UE0-UE1 link resolves to LOS (SNR 45.4-48.1 dB), while UE0-UE2 purposely resolves to NLOSv (SNR 34.6-37.7 dB). The two blocked links (UE1-UE2 in both directions) have a forced path loss of 200 dB and SNR of -78.8 dB, driving CQI to 0, confirming that SET\_LINK\_BLOCK renders the link unusable at the broker. 
This creates a hidden-node topology: UE1 and UE2 can each reach UE0, but not each other, so traffic between them must relay through UE0.
Figure~\ref{fig:multihop_topology} shows the resulting Grafana mesh topology panel, confirming the chain worked as intended: both the UE1-UE0 and UE0-UE2 links measure 150 m with 100\% reception probability, while direct paths between UE1 and UE2 are blocked.

\begin{table}[htbp]
\centering
\vspace{-0.3cm}
\footnotesize
\setlength{\tabcolsep}{4pt}
\caption{Link quality measurements for Line Topology}
\vspace{-0.3cm}
\label{tab:link_quality}
\begin{tabular}{@{}l c r r r c l@{}}
\hline
\textbf{Link} & \textbf{State} & \textbf{Dist.} & \textbf{PL} & \textbf{SNR} & \textbf{CQI} & \textbf{Ch.} \\
 & & (m) & (dB) & (dB) & & \\
\hline
UE0$\to$UE1 & active & 150 & 75.8 & 45.4 & 15 & LOS   \\
UE1$\to$UE0 & active & 150 & 73.1 & 48.1 & 15 & LOS   \\
UE0$\to$UE2 & active & 150 & 83.5 & 37.7 & 15 & NLOSv \\
UE2$\to$UE0 & active & 150 & 86.6 & 34.6 & 15 & NLOSv \\
UE1$\to$UE2 & \textbf{blocked} & 300 & \textbf{200.0} & \textbf{$-78.8$} & \textbf{0} & blk. \\
UE2$\to$UE1 & \textbf{blocked} & 300 & \textbf{200.0} & \textbf{$-78.8$} & \textbf{0} & blk. \\
\hline
\end{tabular}
\vspace{-0.3cm}
\end{table}

Table~\ref{tab:BATMAN_routing_combined} (Line rows) shows the BATMAN-adv routing tables. 
The 3GPP Urban model probabilistically selects between LOS and NLOSv conditions; both are observed across the two active links, consistent with Table~\ref{tab:link_quality}.
The results confirm end-to-end Layer-2 multi-hop routing over a 5G NR sidelink mesh: BATMAN-adv discovers the hidden-node condition and provides relaying hooks at UE0. TQ (Transmission Quality) values in the range 7-47 out of 255 are sufficient for stable routing entries; the variability reflects the probabilistic LOS/NLOSv selection seen in Table~\ref{tab:link_quality}.

\subsection{Multi-hop Mesh Star Topology}
The objective of this scenario is to test the same relaying capability under more demanding conditions, and consider a single central device (a gateway or coordinator) serving several other UEs at once, a pattern common across IoT and edge deployments. We therefore propose a mesh star topology, tested with five UEs: Sync-Ref at the centre, four peripheral UEs at 150 m. Table~\ref{tab:BATMAN_routing_combined} (Star rows) shows the BATMAN-adv routing table: every peripheral UE reaches every other peripheral UE exclusively via Sync-Ref, never through a direct peripheral-to-peripheral path. This enforced the intended centralized topology. TQ values range from 1 to 51 across the ten observed routes, reflecting varying link quality without affecting the enforced routing structure.

\begin{table}[!htb]
\centering
\vspace{-0.2cm}
\small
\caption{BATMAN-adv routing tables: Line, Star, Mobile topologies (SR = Sync-Ref)}
\vspace{-0.3cm}
\label{tab:BATMAN_routing_combined}
\footnotesize
\renewcommand{\arraystretch}{1.15}
\setlength{\tabcolsep}{3pt}
\begin{tabular}{@{}c llll|llll@{}}
\toprule
\textbf{Scen.} & \textbf{Node} & \textbf{Dest.} & \textbf{Next Hop} & \textbf{TQ} &
\textbf{Node} & \textbf{Dest.} & \textbf{Next Hop} & \textbf{TQ} \\
\midrule
\multirow{3}{*}{Line}
 & UE0 & UE1 & UE1 (direct)  & 47 & UE1 & UE2 & UE0 (relay)  & 14 \\
 & UE0 & UE2 & UE2 (direct)  & 37 & UE2 & UE0 & UE0 (direct) & 7  \\
 & UE1 & UE0 & UE0 (direct)  & 13 & UE2 & UE1 & UE0 (relay)  & 27 \\
\midrule
\multirow{6}{*}{Star}
 & UE1 & SR  & direct  & 33 & UE3 & SR  & direct  & 24 \\
 & UE1 & UE2 & via SR  & 10 & UE3 & UE4 & via SR  & 16 \\
 & UE1 & UE4 & via SR  & 1  & UE4 & SR  & direct  & 29 \\
 & UE2 & SR  & direct  & 51 & UE4 & UE1 & via SR  & 11 \\
 & UE2 & UE1 & via SR  & 2  & UE4 & UE3 & via SR  & 15 \\
 & UE2 & UE4 & via SR  & 1  &     &     &         &    \\
\midrule
\multirow{3}{*}{Mobile}
 & UE2 & SR & direct   & 6  & UE3 & UE2  &  via SR  & 21  \\
 & UE2 & UE3  & via SR  & 9  & UE3 & SR  & direct  &  24  \\
 & UE3 &  SR & direct  & 127  &   &  &   &  \\
\bottomrule
\end{tabular}
\vspace{-0.5cm}
\end{table}

\subsection{Multi-hop Mobile Topology}
The objective of this last scenario is to create a moving scenario consisting of two moving UEs,  alongside a fixed Sync-Ref node. All communication remains constrained to the centralised topology, with Sync-Ref as the sole relay node. Mobility is handled by the external mobility module as explained in Section~\ref{sec:mobility_module}. Two mobility profiles are supported: pedestrian (0.5-2 m/s, 3GPP TR 38.901 UMa) and vehicular (15-40 m/s, 3GPP TR 37.885 V2V). Fig.~\ref{fig:topology_all}c shows the Grafana topology view during a run with two moving UEs, confirming the expected connectivity.
 
Because the two mobile UEs continuously change position, their visibility to other nodes fluctuates as links form and break with movement; unlike the static line and star scenarios, no single routing-table snapshot is representative of behaviour over the full run. Table~\ref{tab:BATMAN_routing_combined} (Mobile rows) shows one such snapshot, taken from a single reference UE partway through the experiment. 
The set of reachable destinations varies over time as the RWP trajectory moves each mobile UE in and out of range, but critically, every route observed at every snapshot throughout the run is either a direct link to Sync-Ref or a relayed route via Sync-Ref; no peer-to-peer route via another peripheral UE was recorded at any point. This confirms that the central topology constraint enforced by the propagation module's link-blocking (Section~\ref{sec:propagation_module}) holds even as the underlying link set changes dynamically. The moving UEs also introduce continuous variation in link distance and channel state, visible as SNR fluctuations in the Grafana dashboard, demonstrating mobility-induced channel dynamics on a real 5G NR sidelink stack without any modification to OAI's MAC or PHY layers.

\vspace{-0.2cm}
\subsection{Scalability Analysis}
Scalability was evaluated on two SLICES-RI hardware configurations: a CPU node (Xeon Gold, 8 cores, 8\,GB RAM) and a GPU node (A100 x86\_64, 16 cores, 80\,GB RAM), using the full-mesh P2P topology considered by the reception module. 
\begin{table*}[htbp]
\centering
\small
\caption{Broker resource usage and scheduler status vs.\ number of UEs.}
\vspace{-0.3cm}
\label{tab:scalability}
\begin{tabular}{c c c c c c l}
\hline
\textbf{N UEs} & \textbf{Hardware} & \textbf{Broker CPU} & \textbf{Broker RAM} & \textbf{O($N^2$) Links} & \textbf{ZMQ Conn.} & \textbf{Status} \\
\hline
3  & Xeon 8G  & $\sim$97\% & 135 MB  & 6   & 4  & Stable (PSBCH + PSSCH + BATMAN) \\
7  & Xeon 8G  & --         & --      & 42  & 8  & Stable (PSBCH + PSSCH + BATMAN) \\
15 & Xeon 8G  & 93\%       & 1.06 GB & 210 & 15 & Stable (PSBCH only) \\
17 & Xeon 8G  & 57--91\%   & 1.19 GB & 272 & 17 & Partial (PSBCH only) \\
19 & A100 80G & 3.80\%     & 1.10 GB & 342 & 19 & Partial (PSBCH only) \\
\hline
\end{tabular}%
\vspace{-0.3cm}
\end{table*}

Three stability tiers emerge, as summarised in Table~\ref{tab:scalability}. Up to 7 UEs, the full stack is stable: PSBCH synchronisation, PSSCH data exchange, and BATMAN-adv Layer-2 routing all converge reliably. Beyond 7 UEs, PSSCH and BATMAN-adv become unstable due to the MAC scheduler running out of available slots, but PSBCH synchronisation remains stable up to 15 UEs on the Xeon node and partially up to 17--19 UEs across both platforms.
Three independent software bottlenecks were identified:

\begin{itemize} 
    \item \textbf{ZMQ synchronous blocking:} The single-threaded broker performs $N \times (N-1)$ synchronous ZMQ REQ/REP calls to the propagation module per IQ frame, blocking the event loop. For $N > 15$, new UEs are dropped within the timeout window regardless of idle CPU/GPU capacity.
    \item \textbf{Memory constraints (Circular buffers):}
   While the 80 GB A100 node easily handles the memory footprint, the Xeon node (8 GB) struggled. Each node allocates a 186 MB circular buffer (\texttt{CirSize = 48,880,000} samples) per peer. When attempting to scale to 30 UEs, the Sync-Ref process alone consumes approximately 5.7\,GB of RAM, approaching the hardware limit on standard slots; this configuration did not run reliably, consistent with the connection instability observed beyond 19 UEs in Table~\ref{tab:scalability}.
    \item \textbf{Control/data packet multiplexing:} Beyond $\sim$20 active bidirectional links, the single-threaded TDMA scheduler must multiplex PSSCH between user traffic and BATMAN-adv control packets (OGM), delaying or missing allocations and degrading routing convergence (higher OGM inter-arrival times, lower TQ values).
\end{itemize}

Scalability should however not be measured by the number of UEs, but rather by the number of links. SL-RFSIM currently supports up to 15 UEs and 210 directional links, more than twice the 100 links supported by OAI RFSim (which supports up to 50 UEs). Nevertheless, the broker scalability and SL scheduler will be further improved in future work.

\vspace{-0.3cm}
\section{Conclusion}\label{sec:conclusion}
The growing interest in 5G NR Sidelink for industrial, public safety, and beyond-coverage applications is driving the need for experimental platforms capable of evaluating large-scale multi-hop deployments under realistic wireless conditions. Although OpenAirInterface provides an open-source implementation of the 3GPP SL stack, scalable experimentation of mesh networking has remained largely unexplored.

This paper presented SL-RFSIM, a component-based experimentation framework extending OAI with scalable multi-hop 5G NR SL capabilities. By replacing the conventional RFSim client-server architecture with a broker-based publish/subscribe framework, SL-RFSIM enables arbitrary peer-to-peer connectivity while preserving compatibility with the OAI protocol stack. The framework further integrates modular mobility, propagation, reception and monitoring services, providing a reproducible environment for evaluating wireless mesh networking over 5G NR SL.

Experimental validation demonstrated successful integration of BATMAN-adv Layer-2 routing, realistic channel modelling, and scalable multi-user experimentation on the SLICES-RI infrastructure. The results also identified current scalability bottlenecks, highlighting software limitations that provide useful guidance for future developments of scalable OAI-based NR SL experimentation. Future work will focus on integrating standardised Rel.~19 sidelink relay mechanisms, distributed sidelink scheduling, richer mobility models, and O-RAN-based monitoring interfaces, while extending SL-RFSIM toward larger experimental deployments and additional mesh routing protocols.

\balance
\bibliographystyle{ACM-Reference-Format}
\bibliography{references}

@misc{grafana_docs,
  author = {{Grafana Labs}},
  title = {Introduction to {G}rafana},
  howpublished = {Grafana Documentation},
  url = {https://grafana.com/docs/grafana/latest/introduction/},
  year = {2026},
  note = {Accessed: May 3, 2026}
}

@misc{ionos_influxdb,
  author = {{IONOS}},
  title = {What is {InfluxDB}?},
  howpublished = {IONOS Digital Guide},
  url = {https://www.ionos.com/digitalguide/hosting/technical-matters/what-is-influxdb/},
  year = {2026},
  note = {Accessed: May 3, 2026}
}

@techreport{freespace,
  author = {{International Telecommunication Union (ITU)}},
  title = {Recommendation {ITU-R} {P.525-3}: Calculation of free-space attenuation},
  institution = {ITU},
  year = {2016},
  url = {https://www.itu.int/dms_pubrec/itu-r/rec/p/R-REC-P.525-3-201611-S!!PDF-E.pdf}
}

@techreport{3gpp_tr37885,
  author = {{3GPP}},
  title = {{TR} 37.885: Study on evaluation methodology of new Vehicle-to-Everything ({V2X}) use cases for {LTE} and {NR}},
  institution = {3GPP},
  number = {V15.3.0},
  year = {2019},
}

@techreport{3gpp_tr38901,
  author = {{3GPP/ETSI}},
  title = {{TR} 138 901: 5G; Study on channel model for frequencies from 0.5 to 100 {GHz}},
  institution = {ETSI},
  number = {V16.1.0},
  year = {2020}
}

@inproceedings{poddar2025,
  author = {Poddar, Hitesh and Yoshimura, Tomoki and Ishii, Art},
  title = {Validation of {3GPP} {TR} 38.901 Indoor Hotspot Path Loss Model Based on Measurements Conducted at 6.75, 16.95, 28, and 73 {GHz} for {6G} and Beyond},
  booktitle = {2025 IEEE 101st Vehicular Technology Conference (VTC2025-Spring)},
  address = {Oslo, Norway},
  month = jun,
  year = {2025},
  pages = {1--6}
}

@techreport{c2c_ecc68_2022,
  author = {{Car2Car Communication Consortium}},
  title = {Co-Channel Coexistence Mitigation Methods: Additional Investigation ({WP-2096}) -- Annex A: Path-loss models},
  institution = {C2CCC},
  number = {Version 1.0},
  year = {2022},
  url = {https://www.car-2-car.org/fileadmin/documents/General_Documents/C2CCC_WP_2096_Co-ChannelCoexistence_MitigationMethods_AdditionalInvestigation_V1.0.pdf}
}

@article{kryszkiewicz2024,
  author = {Kryszkiewicz, Pawe{\l} and Kliks, Adrian and Sroka, Pawe{\l} and Sybis, Micha{\l}},
  title = {{V2V} Path Loss Modeling at 26\,{GHz} Based on Real-Traffic Measurements},
  journal = {IEEE Wireless Communications Letters},
  volume = {13},
  number = {12},
  pages = {3633--3637},
  year = {2024},
}

@inproceedings{serunin2022,
  author={Serunin, Igor and Pudeev, Andrey and Hwang, Jin-Yup and Lee, Sang-Wook and Maltsev, Alexander},
  booktitle={2022 13th International Conference on Information and Communication Technology Convergence (ICTC)}, 
  title={An Overview and Performance Analysis of CQI Reporting in 5G NR Sidelink}, 
  year={2022},
  pages={34-39},
}

@techreport{3gpp_ts38214,
  author = {{3GPP}},
  title = {Technical Specification 38.214: {NR}; Physical layer procedures for data},
  institution = {3GPP},
  number = {Release 16},
  year = {2020},
  url = {https://www.3gpp.org/ftp/Specs/archive/38_series/38.214/}
}

@article{lusvarghi2024,
  author = {Lusvarghi, Luca and Coll-Perales, Baldomero and Gozalvez, Javier and Merani, Maria Luisa},
  title = {Link Level Analysis of {NR} {V2X} Sidelink Communications},
  journal = {IEEE Internet of Things Journal},
  volume = {11},
  number = {17},
  pages = {28385--28397},
  year = {2024},
}

@inproceedings{lusvarghi2024vtc,
  author = {Lusvarghi, Luca and Coll-Perales, Baldomero and Gozalvez, Javier},
  title = {Impact of the {DMRS} Pattern on {NR} {V2X} Sidelink Communications},
  booktitle = {IEEE 100th Vehicular Technology Conference (VTC2024-Fall)},
  year = {2024}
}

@inproceedings{morev2x,
  author = {Lusvarghi, Luca and Merani, Maria Luisa},
  title = {{MoReV2X} -- A New Radio Vehicular Communication Module for ns-3},
  booktitle = {2021 IEEE 94th Vehicular Technology Conference (VTC2021-Fall)},
  pages = {1--7},
  year = {2021},
}

@techreport{3gpp_tr36942,
  author = {{ETSI}},
  title = {{LTE}; Evolved Universal Terrestrial Radio Access ({E-UTRA}); Radio Frequency ({RF}) system scenarios},
  institution = {ETSI},
  number = {TR 136 942 V10.3.0},
  year = {2012}
}

@misc{ns3_v2v_pl_src,
  author = {{NS3 Project, SIGNET Lab, University of Padova}},
  title = {{NS3} Source: three-gpp-v2v-propagation-loss-model.cc},
  url = {https://www.nsnam.org/doxygen/d8/de1/three-gpp-v2v-propagation-loss-model_8cc_source.html},
  year = {2026}
}

@misc{ns3_v2v_prob_src,
  author = {{NS3 Project, SIGNET Lab, University of Padova}},
  title = {{NS3} Source: probabilistic-v2v-channel-condition-model.cc},
  url = {https://www.nsnam.org/doxygen/d2/db9/probabilistic-v2v-channel-condition-model_8cc_source.html},
  year = {2026}
}

@article{garciavidal2025mco,
  author = {Garc\'{i}a Vidal, Felipe and Egea L\'{o}pez, Esteban and Santa Lozano, Jos\'{e}},
  title = {Evaluation of multichannel operation mechanisms for vehicular networks},
  journal = {Vehicular Communications},
  volume = {56},
  pages = {100978},
  year = {2025},
}

@article{liu2023comcom,
  author = {Peng Liu and others},
  title = {Towards {5G} new radio sidelink communications: A versatile link-level simulator and performance evaluation},
  journal = {Computer Communications},
  volume = {208},
  pages = {231--243},
  year = {2023},
}

@INPROCEEDINGS{sumo, author = {Pablo Alvarez Lopez and others},  title={Microscopic Traffic Simulation using SUMO},  year={2018}, pages={2575-2582}}

@inproceedings{slices_blueprint,
  author    = {Damien Saucez and Sebastian Gallenm{\"u}ller and
               Raymond Knopp and Nikos Makris and Serge Fdida},
  title     = {Blueprint-based Reproducible Research with the SLICES Research Infrastructure},
  booktitle = {IEEE INFOCOM 2024 Workshops (INFOCOM WKSHPS)},
  year      = {2024}
}

@techreport{qualcomm_sidelink,
  author       = {{Qualcomm Technologies}},
  title        = {NR Sidelink Evolution Toward 5G Advanced},
  institution  = {Qualcomm Technologies Inc.},
  year         = {2023}
}

@techreport{3gpp23287,
  author      = {{3GPP}},
  title       = {Architecture Enhancements for the 5G System (5GS) to Support Vehicle-to-Everything (V2X) Services},
  institution = {3rd Generation Partnership Project},
  number      = {TS 23.287},
  type        = {Technical Specification},
  year        = {2025},
  note        = {Release 19}
}

@techreport{3gpp38836,
  author      = {{3GPP}},
  title       = {Study on NR Sidelink Relay},
  institution = {3rd Generation Partnership Project},
  number      = {TR 38.836},
  type        = {Technical Specification},
  year        = {2025},
  note        = {Release 19}
}

@techreport{5gacia2023,
  author       = {{5G Alliance for Connected Industries and Automation (5G-ACIA)}},
  title        = {Using 5G Sidelink in Industrial Factory Applications},
  institution  = {5G Alliance for Connected Industries and Automation (5G-ACIA)},
  type         = {White Paper},
  month        = jun,
  year         = {2023},
  url          = {https://5g-acia.org/whitepapers/using-5g-sidelink-in-industrial-factory-applications/},
  note         = {Accessed: April 26, 2026}
}

@ARTICLE{lin2025,
  author={Lin, Xingqin},
  journal={IEEE Communications Standards Magazine}, 
  title={The Bridge Toward 6G: 5G-Advanced Evolution in 3GPP Release 19}, 
  year={2025},
  volume={9},
  number={1},
  pages={28-35},
}

@inproceedings{bonati2021colosseum,
  author = {Leonardo Bonati and others},
  title = {Colosseum: Large-Scale Wireless Experimentation Through
           Hardware-in-the-Loop Network Emulation},
  booktitle = {2021 IEEE International Symposium on
               Dynamic Spectrum Access Networks (DySPAN)},
  year = {2021},
  pages = {1--7}
}

@article{kaltenberger2024oai,
  author = {Florian Kaltenberger and others},
  title = {Driving innovation in 6G wireless technologies: The OpenAirInterface approach},
  journal = {Computer Networks, Volume 269, September 2025},
  year = {2025}
}

@article{patriciello2024lena,
  author  = {Giuseppe Piro and Nicola Baldo and Michele Polese and others},
  title   = {The 5G-LENA Simulator: An Open-Source Platform for 5G NR Network Research},
  journal = {Computer Networks},
  year    = {2024}
}

@misc{rfc3561,
  author       = {Charles Perkins and Elizabeth Belding-Royer and Samir Das},
  title        = {Ad hoc On-Demand Distance Vector (AODV) Routing},
  howpublished = {RFC 3561},
  institution  = {IETF},
  year         = {2003},
}

@misc{rfc3626,
  author       = {Thomas Clausen and Philippe Jacquet},
  title        = {Optimized Link State Routing Protocol (OLSR)},
  howpublished = {RFC 3626},
  institution  = {IETF},
  year         = {2003},
}

@misc{batmanadv,
  author       = {{Open-Mesh Project}},
  title        = {B.A.T.M.A.N. Advanced},
  year         = {2025},
  howpublished = {\url{https://www.open-mesh.org/projects/batman-adv}}
}

\end{document}